\documentclass[10pt,letterpaper]{article}
\usepackage[top=0.85in,left=2.75in,footskip=0.75in,marginparwidth=2in]{geometry}

\usepackage[utf8]{inputenc}

\usepackage{cite}

\usepackage{nameref,hyperref}

\usepackage[right]{lineno}

\usepackage{microtype}
\DisableLigatures[f]{encoding = *, family = * }

\raggedright
\usepackage{changepage}

\usepackage[aboveskip=1pt,labelfont=bf,labelsep=period,singlelinecheck=off]{caption}

\makeatletter
\renewcommand{\@biblabel}[1]{\quad#1.}
\makeatother

\usepackage{lastpage,fancyhdr,graphicx}
\usepackage{epstopdf}
\fancyheadoffset[L]{2.25in}
\fancyfootoffset[L]{2.25in}

\usepackage{color}

\definecolor{Gray}{gray}{.25}

\usepackage{graphicx}

\usepackage{sidecap}

\usepackage{wrapfig}
\usepackage[pscoord]{eso-pic}
\usepackage[fulladjust]{marginnote}
\reversemarginpar

\begin{document}
\vspace*{0.35in}

\begin{flushleft}
{\Large
\textbf\newline{A biomimetic kidney tubule model.}
}
\newline
\\
Elod Mehes \textsuperscript{1},
Tana S Pottorf \textsuperscript{2},
Marton Gulyas \textsuperscript{1},
Sandor Paku \textsuperscript{3},
Pamela V. Tran \textsuperscript{2},
Andras Czirok \textsuperscript{1,2,*}
\\
\bigskip
\bf{1} Dept of Biological Physics, Eotvos University, Budapest, Hungary
\\
\bf{2} Dept of Anatomy \& Cell Biology and the Jared Grantham Kidney Institute, University of Kansas Medical Center, Kansas City, USA
\\
\bf{3} First Department of Pathology and Experimental Cancer Research, Semmelweis University, Budapest, Hungary
\\
\bigskip

correspondence: * aczirok@gmail.com

\end{flushleft}

\section*{Abstract}

A critical barrier in the nephrology field is the lack of appropriate {\em in vitro} renal tubule models that allow manipulation of various mechanical factors, facilitating studies of disease pathophysiology and drug discovery. Here we report development of a novel {\em in vitro} assay system comprised of a renal tubule within an elasto-plastic extracellular matrix microenvironment. This {\em in vitro} tubule mimetic device consists of a container with two, pipette-accessible ports, filament-deposition (3D-) printed into 35 mm cell culture dishes. The container is filled with a hydrogel, such as a collagen I or fibrin gel, while a narrow masking tube is threaded through the ports. Following gelation, the masking material is pulled out leaving a tunnel within the gel. Seeding of the tunnels with M1 or MDCK renal epithelial cells through the side ports results in a monolayer with apical-basal polarity, such that laminin and fibronectin are present on the basal surface, while primary cilia project from the apical side of cells into the tubular lumen. The device is optically accessible, and can be live-imaged by phase contrast or epifluorescence microscopy. The lumen of the epithelial-lined tube can be connected through the side ports to a circulatory flow. We demonstrate that kidney epithelial cells are able to adjust the diameter of the model tubule by myosin-II dependent contractility. Furthermore, cells of the tubule are also able to remodel the surrounding hydrogel leading to budding from the main tubule. We propose that this versatile {\em in vitro} model system can be developed into a future pre-clinical tool to study pathophysiology of kidney diseases and identify therapeutic compounds.



\section{Introduction}
The kidneys are vital organs that perform multiple functions, including blood filtration, osmoregulation, blood pressure and blood pH regulation, and Vitamin D activation. The renal parenchyma is organized into more than a million nephrons, each of which begins with a glomerulus that filters blood, and concomitantly, forms a filtrate. Subsequently, this filtrate passes through various segments of a tubule lined with renal epithelial cells, which selectively reabsorb and secrete nutrients, ions and wastes, modifying the intraluminal filtrate. As the fluid passes through the tubular lumen, the flow exerts hydrostatic pressure and shear stress at the apical surface of epithelial cells lining the renal tubule. Each cell is equipped with the appropriate cellular structure and molecular machinery to respond to the mechanical forces within the tissue microenvironment \cite{pmid12920399, pmid15972389, pmid31363000}. Other mechanical factors that impinge on a renal epithelial cell include stiffness of the underlying basement membrane, and connection to adjacent cells. Faulty mechanosensing, i.e. inadequate response to these mechanical factors has been implicated in kidney disease, in particular, Autosomal Dominant Polycystic Kidney Disease (ADPKD), a genetic disease which causes progressive growth of fluid-filled renal cysts \cite{pmid29326913, pmid29600752, pmid26146837}.

An improved understanding of how mechanical factors of the tissue microenvironment contribute to renal tubule homeostasis and disease is still hindered by the lack of appropriate {\it in vitro} models. An epithelial monolayer within a flow chamber allows modulation of fluid flow \cite{pmid26146837}, however, these cells cannot undergo morphogenetic processes. Pluripotent or adult stem cells can be programmed to form a kidney with diverse cell types, but these organoid models cannot be readily subjected to hydrodynamic stress \cite{pmid24009235}. More recent organ-on-a-chip platforms allow formation of renal tubules to which intraluminal flow can be applied using a rocker, but precise calculation of the flow and pressure cannot be achieved, and the chemically crosslinked nature of the extracellular matrix (ECM) environment precludes morphogenesis \cite{pmid30833775}. The ideal {\em in vitro} model should allow manipulation of tissue mechanical factors like apical fluid pressure, shear stress, basement membrane stiffness, tissue elasticity and contractility, as well as the evaluation of environment-induced cell responses. Furthermore, the experimental system should enable the epithelial cells to also modulate the mechanical state of the tissue environment, thereby creating feedback regulation.  

Here we report a 3D-printed {\em in vitro} 3D kidney tubule-mimetic device that allows systematic regulation of intraluminal flow and examination of various mechanical factors in totality. With the ability to connect this {\em in vitro} culture system to live cell imaging microscopy, we show that the tubules are contractile and remodel the surrounding ECM environment.

\section{Materials and methods}

\subsection{Cell cultures}
MDCK (Madin-Darby canine kidney) cells were purchased from ATCC and were cultured in DMEM medium (Lonza, 12-604F) supplemented with 10\% FBS (Gibco). M1 mouse renal cortical collecting duct epithelial cells \cite{pmid1654478} were obtained from Dr James Calvet (University of Kansas Medical Center) and cultured in DMEM/F12 medium (Bio Sera, LM-D1223) supplemented with 10\% FBS (Gibco). All cultures were maintained in 6-well culture plates (Greiner) at 37$^o$C in a humidified incubator with 5\% CO$_2$ atmosphere. Once confluent, cell monolayers were washed in phosphate-buffered saline (PBS) twice, briefly incubated in trypsin-EDTA (Lonza), then resuspended in their respective culture medium as a cell suspension. These suspensions were used to seed the tubule mimetic device.

\subsection{Hydrogels}
Type I collagen gels were prepared with 3.0 mg/ml collagen concentration (rat tail collagen I, Corning) according to the manufacturer’s gel preparation protocol. Briefly, collagen I dissolved in 20 mM acetic acid was neutralized with 1M NaOH and 10x PBS (Lonza) and transferred to 37$^o$C. Gelation of collagen was completed in the mimetic device chamber in 30 minutes. 

Fibrin gels with 3 mg/ml fibrin concentration were prepared as described earlier \cite{pmid16249343}. Briefly, 3 mg/ml human fibrinogen was combined with 200 U/ml aprotinin, 2 U/ml human thrombin, 2.5 mM CaCl$_2$ (all from Sigma) and 2 U/ml human factor XIII (CSL Behring) then the solution was transferred and allowed to gelate in the flow chamber arena.

\subsection{Flow chamber fabrication}
Hydrogel chambers for the kidney tubule mimetic device were filament-deposition (3D-) printed using commercial polylactic acid (PLA, Verbatim) into 35 mm cell culture dishes (Greiner) as described in detail in \cite{pmid30180178}. The hydrogel chamber is an open round arena with 6 mm diameter and 50 $\mu$l volume.  We threaded either a teflon tube of 500 $\mu$m outer diameter (Sailu Technology) or a damyl line of 250 micron diameter (Titan) through the opposite ports of the PLA structure -- serving as masks for the tunnel spanning the hydrogel chamber. After sterilization with 70\% ethanol, the arena is filled with a hydrogel solution. After gelation the masking tube or line is removed leaving a 6 mm long void tunnel connecting the two open ports. The flow chamber is then filled up with cell culture medium and placed in the CO2 incubator at 37$^o$C for 2 hours to let the hydrogel equilibrate with medium. After equilibration, the tunnel is seeded with renal epithelial cells injected through either port as a dense cell suspension (10 000 cell/microlitre) readily prepared from a monolayer culture by washing up cells after brief incubation with trypsin-EDTA solution (Sigma). Cell attachment, proliferation and migration on the hydrogel tunnel wall eventually results in full lining of the entire surface of the tunnel by a monolayer of cells in 1 to 3 days under standard cell culture conditions.

\subsection{Medium circulation} 
Once a cell monolayer is established in the tunnels, the kidney tubule mimetic devices can be connected to circulation through the side ports of the hydrogel chamber. While culture medium covers the entire chamber in the culture dish, the circulatory flow of the culture medium through the tunnel is driven by a peristaltic pump (Ismatech, Mini-S 820) equipped with a cell culture grade peristaltic pump tube (Cole-Parmer, Tygon E-3603, ID=1.4 mm). We used teflon tubing (Bola, S1810-12) of 1 mm inner diameter to connect the pump and the devices within a microscope stage-top incubator. The typical flow rate of the pump was set to 0.66 $\mu$l/s.

\subsection{Histology and immunolabeling }
For histology and immunolabeling, {\em in vitro} tubules were fixed with 4\% paraformaldehyde  in PBS. Some renal tubules were embedded in paraffin, and sectioned at 7$\mu$m thicknesses using a microtome. For histology, sections were deparaffinized, rehydrated through a series of ethanol washes and stained with hematoxylin and eosin (H\&E) or toluidine blue. For immunodetection of cilia, renal tubule sections were deparaffinized, rehydrated, then subjected to antigen retrieval. Tissue sections were steamed for 15 minutes in Sodium Citrate Buffer (10 mM Sodium Citrate, 0.05\% Tween 20, pH 6.0), returned to room temperature, rinsed 10 times in distilled water, and washed 5 minutes in PBS. Sections were blocked with 1\% BSA in PBS for 1 hour at room temperature, and incubated with primary antibodies against ARL13B (1:300; Proteintech) or acetylated $\alpha$-tubulin (1:4000; Sigma), overnight at 4$^o$C. Sections were washed three times in PBS, then incubated with secondary antibodies conjugated to AlexaFluor 594 (1:500; Invitrogen by Thermo Fisher Scientific) for 1 hour at room temperature. Sections were washed three times in PBS, then mounted with Fluoromount G with DAPI.

For immunodetection of laminin, rabbit polyclonal anti-laminin (Dako, Z0097) was used at 1:200 dilution overnight at 4$^o$C as primary antibody and anti-rabbit Ig-Alexa555 (Life Technologies) was used as secondary antibody at 1:200 dilution for 3 hours at room temperature. After immunolabeling in wholemount conditions, the samples were embedded in Cryomatrix resin (Thermo Scientific) and sectioned using a Cryostar NX50 cryostat (Thermo Scientific) to obtain 14 $\mu$m sections which were mounted on microscopic slides (Thermo Scientific) in a mounting medium with NucBlue counterstain (Thermo Fisher) and imaged subsequently.

\subsection{Brightfield and epifluorescent microscopy}
Fluorescent imaging of the {\em in vitro} renal epithelial tissue was performed on a Zeiss Axio Observer Z1 inverted microscope with 40x EC Plan-Neofluar objective and Zeiss AxioCam MRm CCD camera. Alternatively, immuno- and histological stained sections were imaged using a Nikon 80i microscope with a Nikon DS-Fi1 camera.

\subsection{Transmission electron microscopy}
Renal epithelial tubules embedded in collagen or fibrin gel, were fixed in 4\% paraformaldehyde/2\% glutaraldehyde, then post-fixed in osmium tetroxide. Samples were dehydrated in a graded series of ethanol and propylene oxide, and embedded in EMbed 812 resin. Semi-thin sections were cut with a diamond histo knife, stained with toluidine blue, and selected for thinning. Thin sections were cut with a diamond histo knife and placed on copper grids that were then stained with uranyl acetate and lead citrate. Samples were viewed and imaged using a JEOL JEM-1400 transmission electron microscope equipped with a Lab6 gun. 

\subsection{Scanning electron microscopy}
Samples were fixed in 4\% paraformaldehyde/2\% glutaraldehyde, then washed 3 times (10 minutes each) with 0.1 M Na-cacodylate, pH 7.4. Samples were post-fixed for 30 minutes with 1\% OsO4 in 0.1 M Na-cacodylate buffer. Samples were dehydrated using an ethanol series, followed wth hexamethyldisilazane (HMDS; Electron Microscopy Sciences). Samples were mounted onto metal stubs and sputter coated with gold. Samples were viewed and imaged using a Hitachi S-2700 Scanning Electron Microscope equipped with a Quartz PCI digital capture.

\subsection{Live cell imaging}
Time-lapse recordings of kidney epithelial cells in hydrogel tunnels were performed on a Zeiss Axio Observer Z1 inverted microscope with 10x Plan Neofluar objective. The microscope was equipped with a Zeiss AxioCam MRm CCD camera and a Marzhauser SCAN-IM powered stage. Cell cultures established in flow chambers within cell culture dishes (Greiner) were kept in a stage-mounted incubator (Cell Movie) providing 37$^o$C and a humidified 5\% CO$_2$ atmosphere. Stage positioning, focusing and image collection were controlled by Zeiss Axiovision 4.8 software and a custom experiment manager software module. Phase contrast images were collected every 10 minutes from several microscopic fields for up to 4 days.

\subsection{Contraction analysis}

Contraction of renal tubule models were analyzed on the basis of image series captured by phase-contrast time-lapse videomicroscopy. Using NIH ImageJ software’s Reslice function image series were resliced along a cross sectional line across the tubule resulting in a kymogram with spatial and temporal calibration. Contraction or dilation of the tubule was read from the calibrated kymograms.

\subsection{Cell contractility inhibitor}
For inhibition of actomyosin contractility, Y27632, a cell-permeable Rho kinase inhibitor, (Merck Millipore) was applied. The inhibitor was dissolved in water and kept in 10~mM stock solutions and used at 50 $\mu$M final concentration. For control treatments, an identical volume of water was applied.

\section{Results}

\subsection{Kidney tubule model device}

\begin{figure}
\begin{center}
\includegraphics[width=4in]{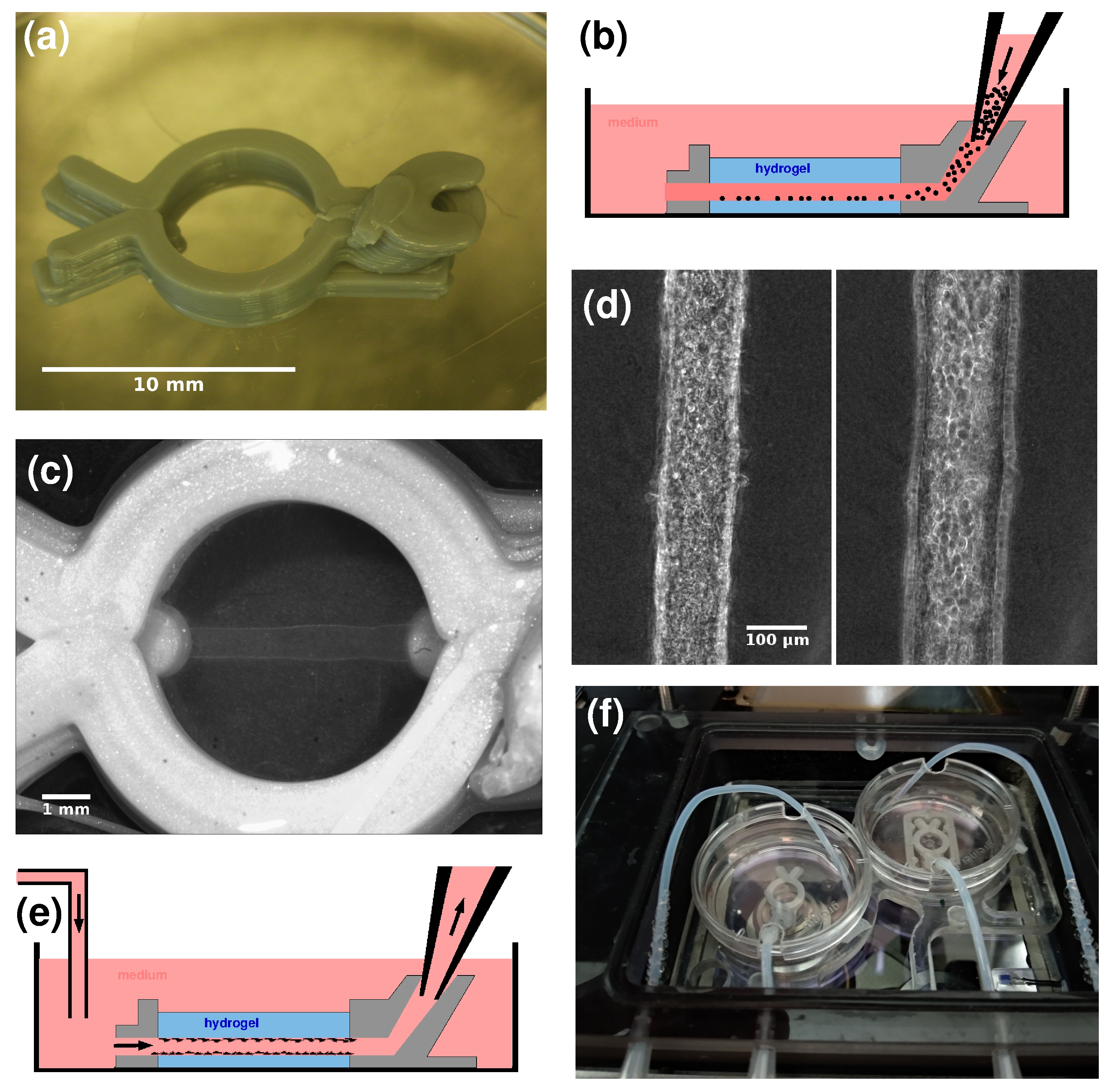}
\caption{\small
Flow chamber device for renal tubular model. (a): Flow chamber device 3D-printed in a cell culture dish, scale bar: 10 mm. (b): Schematic diagram of cell seeding into the hydrogel tunnel of flow chamber using a pipette tip at one of the chamber's ports. (c): Hydrogel tunnel in the flow chamber, scale bar: 1 mm; (d): Phase-contrast images of the bottom (left) and top (right) of a tunnel lined by M1 renal epithelial cells. Scale bar: 100 $\mu$m. (e): Schematic diagram of medium circulation in the kidney tubule mimetic device. (f): Two flow chambers placed within a microscope stage-top incubator.
}
\label{device} 
\end{center}
\end{figure}

Our tubular flow chamber device is a polylactic acid (PLA) structure consisting of a round container arena and two opposite ports, 3D-printed into a 35 mm cell culture dish (Fig.~\ref{device}(a)). The arena is filled with a hydrogel substance surrounding a 6 mm long and 200-500 $\mu$m diameter tunnel mask that spans the arena and connects the two ports (Fig.~\ref{device}(c)). The hydrogel can be a type I collagen gel, a fibrin gel, or other combinations of extracellular matrix (ECM) proteins that are not presented here. The mechanical characteristics of the hydrogel are determined by the ECM protein type and its concentration. After removing the masking material, the hydrogel tunnel is submerged in cell culture medium and  seeded with renal epithelial cells (Fig.~\ref{device}(b)). The structure is kept under standard cell culture conditions, whereby epithelial cells migrate and proliferate within the tunnel, covering its surface as a cell monolayer. Devices can be placed in a microscope stage-top cell culture incubator and subjected to live imaging. The morphology and motility of the cells, as well as changes in the size and shape of the renal tubule model can be observed and recorded with live imaging microscopy for several days (Fig.~\ref{device}(d)-(f)). 

Cells lining the tunnel wall can be exposed to hydrodynamic shear forces by connecting the device to a circulatory flow (Fig.~\ref{device}(e)): a peristaltic pump pulls the medium from one of the ports, and returns to the reservoir inside the culture dish. Thus the flow of medium in the renal tubule model is controlled by the rotational speed of a peristaltic pump. As an example, the peristaltic pump flow rate of $\phi=0.66$ $\mu\ell$/s yields a flow speed of $v=1.3$ mm/s in the tubule. If the viscosity of the medium is $\eta\approx 1$ mPa, the corresponding shear stress at the walls of a $d=650$ $\mu$m wide tunnel is $\tau = 32\eta\phi/d^3\pi \approx 24$ mPa. In comparison, a collecting duct tubule diameter is $d=20–50$ $\mu$m, and a $\phi\approx 1$ $\mu\ell$/h filtrate flow rate translates to an average flow velocity of $v \approx 4\phi/d^2\pi \approx 0.6 - 3$ mm/sec and a corresponding wall shear stress of $\tau \approx 20-300$ mPa. Thus, flow velocities and wall shear stresses in the kidney tubule model are within the physiological range.

\subsection{Epithelial morphology}

\begin{figure}
\begin{center}
\includegraphics[width=4in]{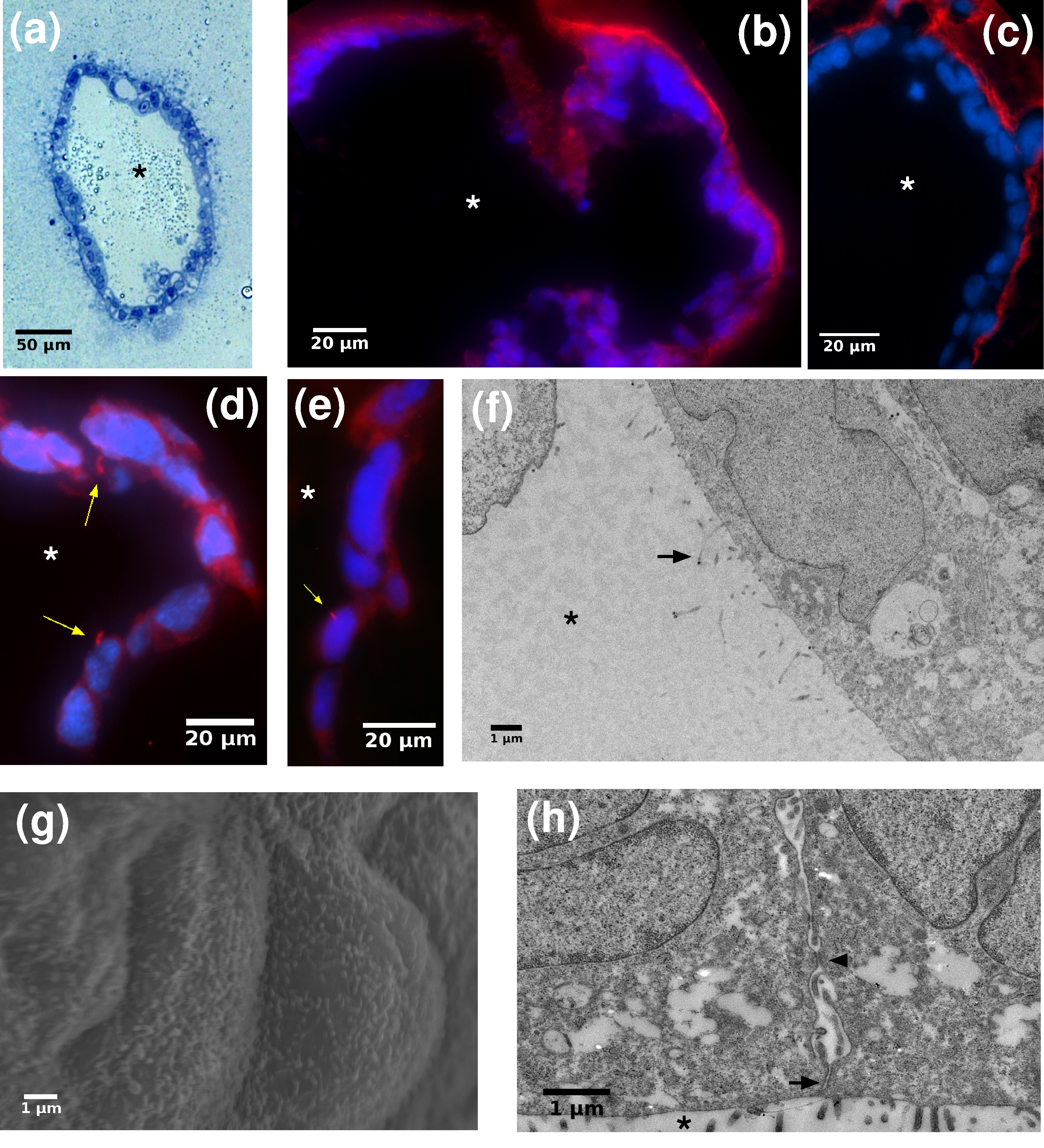}
\caption{\small
Histology of the renal tubule model. (a): Cross section of MDCK tubule in collagen I gel matrix, toluidine blue staining. (b)-(c): Immunofluorescence labeling (red) of basement membrane proteins in tubules of M1 cells: laminin (b) and fibronectin (c). (d)-(e): Immunolocalization (red) of primary cilia on M1 cells (arrows) -- acetylated alpha-tubulin (d), Arl13B (e). (f): TEM image of M1 cells, arrow points to microvilli. (g): SEM image shows doming of MDCK cells and microvilli on the apical surface. (h): TEM image of MDCK cells, showing tight junctions (arrow) and desmosomes (arrowhead).  Scale bars: 50 $\mu$m in (a), 20 $\mu$m in (b)-(e), 1 $\mu$m in (f)-(h). Asterisks indicate the tubule lumen in (a)-(h), nuclei are DAPI-labelled (blue) in (b)-(e).
}
\label{histology} 
\end{center}
\end{figure}

We seeded collagen I and fibrin gel tunnels with either M1 mouse cortical collecting duct cells or MDCK canine distal tubule cells. After two weeks in culture, we characterized the epithelium layer that developed in the kidney tubule mimetic. Histology and toluidine blue staining of semi-thin sections revealed a cell monolayer with a basement membrane (Fig.~\ref{histology}(a)-(c); data not shown). Immunolabeling of sections revealed that the basement membrane was rich in laminin and fibronectin. These ECM components were secreted and assembled by the renal epithelial cells on the basal side of cells. On the apical surface projecting into the tubular lumen, epithelial cells sprouted a primary cilium, which was immunostained for acetylated alpha-tubulin, marking the microtubular axoneme, or for Arl13B, marking the ciliary membrane(Fig.~\ref{histology}(d)-(e)). Transmission electron microscopy (TEM) showed the presence of microvilli also on the apical surface of cells (Fig.~\ref{histology}(f)). Scanning electron microscopy (SEM) showed microvilli as well as doming of the apical surface of renal epithelial cells ((Fig.~\ref{histology}(g)). Finally, TEM showed tight junctions connecting adjacent cells (Fig.~\ref{histology}(h)). These data indicate that in the mimetic tubes, renal epithelial cells form a monolayer with proper epithelial polarity and morphology.

\subsection{Epithelium is in a contractile state}

\begin{figure}
\begin{center}
\includegraphics[width=4in]{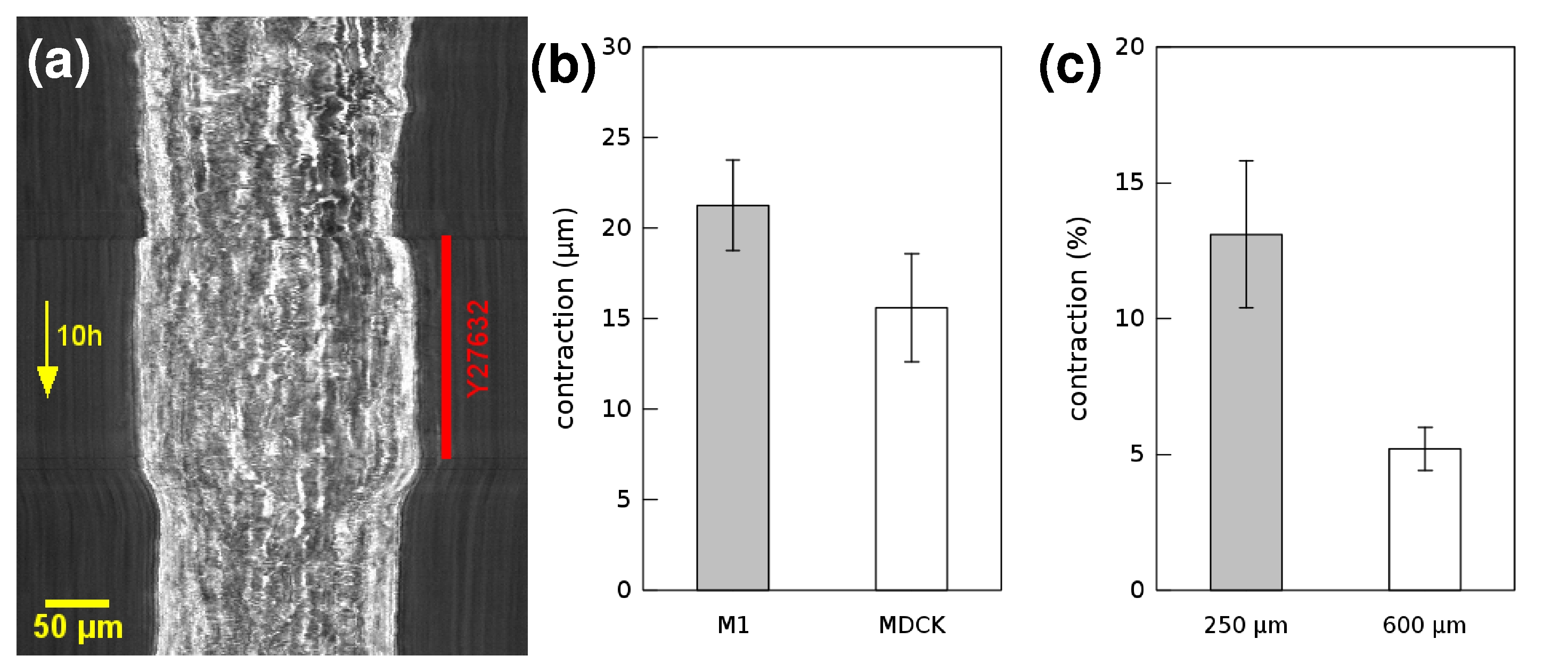}
\caption{\small
Contractility of model renal tubules. (a): A representative kymogram along a perpendicular cross section of a tubule of M1 cells in collagen I gel matrix. Time progresses from top to bottom, a 10h long interval is indicated by the yellow arrow. The duration of treatment with 50 $\mu$M Y27632 Rho kinase inhibitor is indicated with the red line. Scale bar: 50 $\mu$m. (b)-(c): Quantitative analysis of tubule contractility. (b): Contraction of 250 $\mu$m diameter tubules of M1 or MDCK cells, embedded in collagen I gels. Error bars indicate SEM, n=2. (c): Contraction of thin (250 $\mu$m) or wide (600 $\mu$m) tubules of M1 cells, expressed as percent of untreated tubule diameter. Error bars represent SEM.
}
\label{contraction} 
\end{center}
\end{figure}

To study the contractility of our mimetic renal tubule epithelium, we used the Rho kinase inhibitor Y27632 to reversibly inhibit actomyosin contractile function. The inhibitor was applied at 50 $\mu$M final concentration in the flow chamber without circulation, resulting in dilation of the tunnel within 30 minutes. Subsequent washout of the inhibitor caused a gradual contraction of the tunnel to approximately its initial state. Both the dilation and contraction responses were quantitatively analyzed by constructing kymograms from the phase-contrast time-lapse image series (Fig.~\ref{contraction}(a)). In a collagen I gel, the diameter of 250 $\mu$m wide M1- or MDCK-lined tunnels contracted by 20 or 15 $\mu$m, respectively (Fig.~\ref{contraction}(b)). The relative contraction (as defined by the ratio of the measured contraction to tubule diameter) is larger for tubules with smaller diameters: while the 250 $\mu$m tubule contracts almost 13\% of its diameter, a 600 $\mu$m large tubule contracts only about 5\%  (Fig.~\ref{contraction}(c)).

\subsection{Tubule remodeling}

\begin{figure}
\begin{center}
\includegraphics[width=4in]{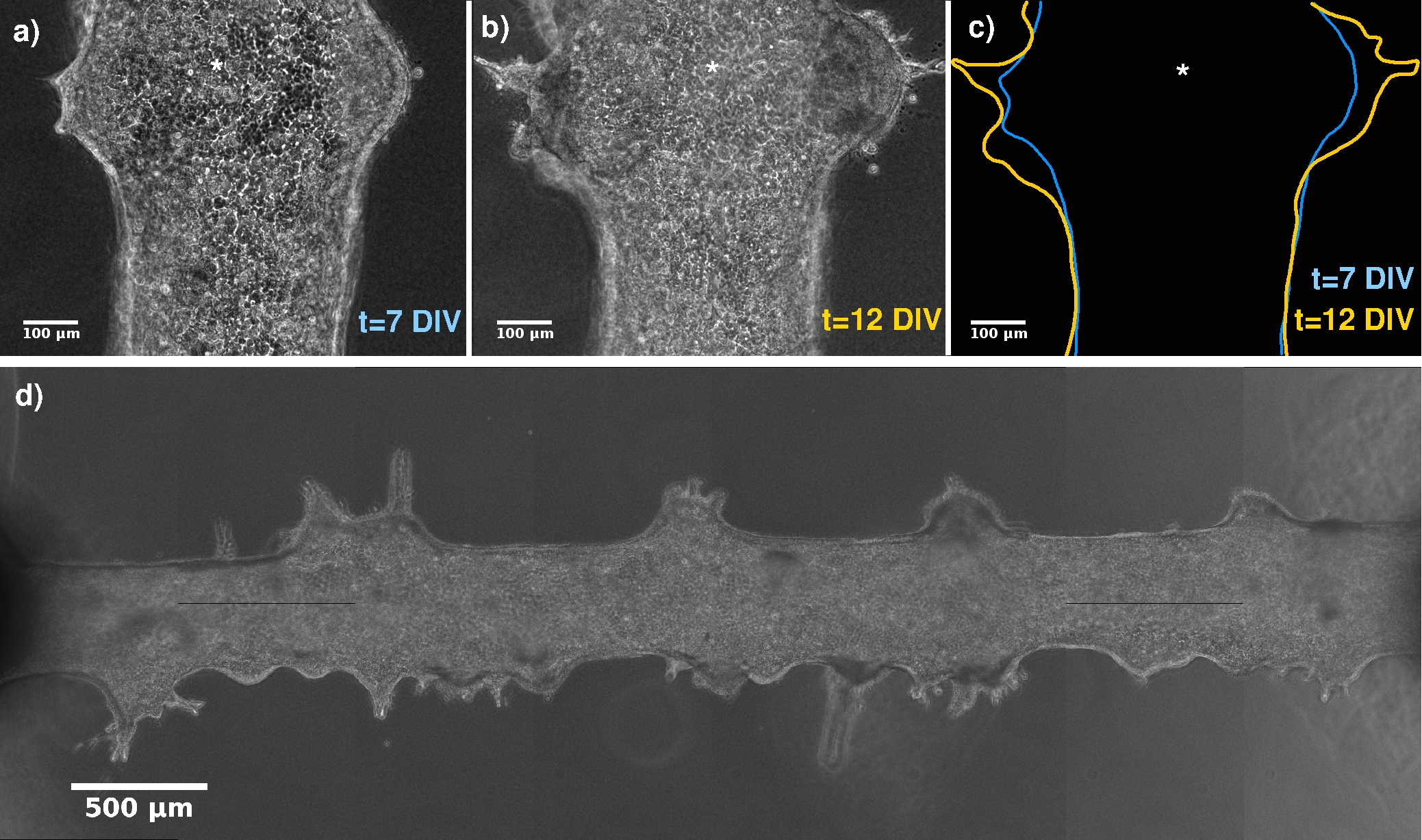}
\caption{\small
Renal tubules can remodel the hydrogel and thereby change morphology. (a)-(b): A tubule of M1 cells in a collagen I gel exhibits bud extensions, at DIV 7 (a) and DIV 12 (b).  (c): Comparison of tubule wall outlines at DIV 7 (blue lines) and DIV 12 (orange lines). Asterisks indicate tubule lumen, scale bar: 100 $\mu$m. (d): Phase-contrast image of the full length tubule at DIV 16, scale bar: 500 $\mu$m.
}
\label{remodeling} 
\end{center}
\end{figure}

The tubules can be maintained in the mimetic devices for 2-3 weeks. During this time frame the kidney epithelial cells interact with and remodel the surrounding hydrogel. Interestingly, tubules can form buds -- sphere-like extensions along the tunnel wall, as cells slowly expand into the surrounding ECM (Fig.~\ref{remodeling}). The location of buds likely corresponds to softer hydrogel patches. Within such areas epithelial cells are also often observed invading the hydrogel in multicellular sprouts. Thus the kidney tubule mimetic device has profound potential to study epithelial patterning, including cyst formation, and its guidance by tissue mechanical factors.

\section{Discussion}

The best {\em in vitro} models allow exploration of questions not possible to answer using {\em in vivo} models, and thus, are essential to deepening our understanding of mechanisms underlying normal tissue development, homeostasis and disease pathogenesis. For example, {\it in vitro} models of ADPKD have been instrumental in identifying molecular mechanisms and screening of pharmacological compounds. These models include an embryonic metanephric organ culture, which involves culturing mid-gestational kidneys isolated from mouse models \cite{pmid17108316, pmid31395378}, growth of {\it in vitro} microcysts using ADPKD patient-derived renal epithelial cells cultured in a collagen gel with growth factors \cite{pmid31395386}, and programming of patient-derived induced pluripotent stem (iPS) cells into kidney organoids \cite{pmid26493500}. While the embryonic metanephric organ culture and iPS-derived organoids model a developing kidney -- outside the ADPKD field -- the use of adult stem cells to generate kidney organoids, and culturing of renal epithelial cells on an organ-on-a-chip platform, have been reported to model the adult kidney and adult renal tubule, respectively \cite{pmid32466429,pmid30833775}.

Since the kidney is comprised of renal tubules through which filtrate passes, renal epithelial cells are subjected to intraluminal flow and pressure. Of the models described above, only the {\em in vitro} renal tubules generated on an organ-on-a-chip platform allow for intraluminal flow \cite{pmid30833775, pmid32401606}. Since microfluidic and micromanufacturing capacities are often limited, our 3D printing technique offers a simpler, readily usable approach, easily scaled up to several parallel cultures.  In addition, we have coupled our {\em in vitro} system to live microscopy, enabling the renal epithelial cells to be viewed at any time point. Our system is extremely versatile. Similar to the LumeNEXT system \cite{pmid26610188}, which has been used to create {\em in vitro} models of blood vessels, lymph vessels, and mammary ducts, our model can also be used to create {\em in vitro} blood vessels (data not shown), and other biological systems that have a lumen, such as the biliary system. 

Our data demonstrate that renal epithelial cells seeded in the tunnel form a monolayer, with proper apical-basal polarity and formation of tight junctions between adjacent cells. Additionally, the {\em in vitro} renal tubule is able to dilate and contract within the extracellular matrix, in response to changes in ROCK signaling. Mechanical factors are an important component in regulating tissue development and homeostasis, and in turn, disease pathogenesis. Yet such studies in nephrology have been limited due to the lack of appropriate models. Our system provides a novel {\em in vitro} 3D renal tubular model coupled to a continuous optical monitoring system that is extremely versatile, and  for the first time, enables systematic study of the role of intraluminal flow and pressure, and of other mechanical factors in totality in kidney homeostasis and disease.

\section*{Acknowledgments}
We thank members of the Department of Anatomy and Cell Biology and of the Jared
Grantham Kidney Institute for helpful discussions. We thank Pat St. John and
Larysa Stroganova of the KUMC Electron Microscopy Research Laboratory, which is
supported by COBRE grant P20GM104936.  We also thank Jing Huang of the KUMC
Histology Core, which is supported by NIH U54HD090216 and COBRE NIH
P30GM122731. This work was also supported by a PKD/KU Cancer Center pilot grant (P30 DK106912 to AC and PVT), the NIH (R01DK103033 to PVT), and 
the Hungarian National Research, Development and Innovation Office (OTKA-FWF ANN 132225 to AC).  

\nolinenumbers


\bibliographystyle{abbrv}

\end{document}